\documentstyle{article}

\begin{document}



\font\tt=cmtex10 at 10pt
\font\tts=cmtex10 at 8pt
\font\cal=cmsy10 at 10pt
\font\cals=cmsy10 at 8pt
\font\itb=cmbxsl10 at 10pt

\font\boldrom=cmb10 at 10pt

\font\bigboldrom=cmb10 at 12pt
\font\biggerboldrom=cmb10 at 15pt
\font\brm=cmb10 at 10pt

\font\scriptsize=cmr8 at 8pt

\font\PART=cmr17 at 17pt

\font\SS=cmss10 at 10pt
\font\SSs=cmss10 at 8pt
\font\SSf=cmssbx10
\font\SSp=cmss10 at 9pt

\def\sect{\S}

\def\a{\alpha}
\def\A{\Alpha}
\def\Alpha{\Arm}
\def\b{\beta}
\def\B{{\Beta}}

\def\Chi{\Xrm}

\def\D{\Delta}
\def\d{\delta}

\def\e{\epsilon}

\def\F{\Phi}
\def\f{\phi}
\def\G{\Gamma}
\def\g{\gamma}

\def\i{\iota}

\def\vf{\varphi}
\def\k{\kappa}

\def\L{\Lambda}
\def\l{\lambda}

\def\M{\Mu}
\def\Mu{\Mrm}
\def\m{\mu}
\def\n{\nu}
\def\nab{\nabla}
\def\N{\Nu}
\def\Nu{\Nrm}

\def\p{\pi}
\def\vr{\varrho}
\def\r{\rho}
\def\Si{\Sigma}
\def\si{\sigma}

\def\s{\sigma}
\def\t{\tau}
\def\Th{\Theta}
\def\th{\theta}

\def\W{\Om}
\def\w{\omega}
\def\Y{\Psi}
\def\y{\psi}
\def\z{\zeta}


\def\nb{\bf n}
\def\Sb{{\bf S}}


\def\CSSs{{\hbox{\SSs C}}} 
\def\CSS{\hbox{\SS C}}     
\def\ESS{\hbox{\SS E}}
\def\FSS{\hbox{\SS F}}
\def\Gss{\hbox{\SSs G}}
\def\GSS{\hbox{\SS G}}
\def\gSS{\hbox{\SS g}}
\def\gss{\hbox{\SSs g}}

\def\hSS{\hbox{\SS h}}

\def\Hss{{\hbox{\SSs H}}}
\def\HSS{\hbox{\SS H}}
\def\ISS{\hbox{\SS I}}
\def\Mss{{\hbox{\SSs M}}}
\def\MSS{\hbox{\SS M}}

\def\NSS{\hbox{\SS N}}
\def\RSS{\hbox{\SS R}}
\def\tSS{\hbox{\SS t}}
\def\Tss{{\hbox{\SSs T}}}
\def\TSS{\hbox{\SS T}}
\def\TSSp{\hbox{\SSp T}}

\def\VSS{\hbox{\SS V}}
\def\XSS{\hbox{\SSf X}}
\def\ZSS{\hbox{\SS Z}}


\def\Cbb{\ISS{}\kern-4pt\CSS}
\def\Nbb{\ISS{}\kern-1pt{}\NSS}
\def\Rbb{\ISS{}\kern-1pt{}\RSS}
\def\Zbb{\ZSS{}\kern-4pt\ZSS}


\def\c{\cSS}
\def\hbar{\hSS\kern-4.8pt{}\rule[5.2pt]{3pt}{.5pt}}
\def\tav{
			{\hbox{\kern2pt
			\rule[0pt]{3pt}{.5pt}{\kern-3.5pt}
			\rule[0pt]{.5pt}{6pt}{\kern-3.6pt}
			\rule[6pt]{4pt}{.5pt}{\kern-3.6pt}
			\rule[0pt]{.5pt}{6pt}
			}}}
\def\tbar{\tav}


\def\calA{\hbox{\cal A}}
\def\calB{\hbox{\cal B}}
\def\calC{\hbox{\cal C}}
\def\calD{\hbox{\cal D}}
\def\calF{\hbox{\cal F}}
\def\calH{\hbox{\cal H}}
\def\calL{\hbox{\cal L}}
\def\calM{\hbox{\cal M}}
\def\calN{\hbox{\cal N}}
\def\calP{\hbox{\cal P}}

\def\calQ{\hbox{\cal Q}}
\def\calR{\hbox{\cal R}}
\def\calS{\hbox{\cal S}}

\def\calT{\hbox{\cal T}}
\def\calV{\hbox{\cal V}}
\def\calX{\hbox{\cal X}}
\def\calZ{\hbox{\cal Z}}

\def\tta{\hbox{\tt a}}
\def\ttA{\hbox{\tt A}}
\def\ttb{\hbox{\tt b}}
\def\ttB{\hbox{\tt B}}
\def\ttC{\hbox{\tt C}}
\def\ttD{\hbox{\tt D}}
\def\ttF{\hbox{\tt F}}
\def\ttH{\hbox{\tt H}}
\def\ttm{\hbox{\tt m}}
\def\ttM{\hbox{\tt M}}
\def\ttN{\hbox{\tt N}}
\def\ttn{\hbox{\tt n}}
\def\ttp{\hbox{\tt p}}
\def\ttQ{\hbox{\tt Q}}
\def\ttR{\hbox{\tt R}}
\def\ttS{\hbox{\tt S}}
\def\ttT{\hbox{\tt T}}
\def\ttV{\hbox{\tt V}}

\def\ttsc{{}^{\hbox{\tts c}}}
\def\ttsq{{}^{\hbox{\tts q}}}
\def\ttsM{{}^{\hbox{\tts M}}}
\def\ttsN{{}^{\hbox{\tts N}}}
\def\ttsq{{}^{\hbox{\tts q}}}
\def\ttsT{{}^{\hbox{\tts T}}}
\def\ttsV{{}^{\hbox{\tts V}}}


\def\Arm{{\rm A}}
\def\Arm{{\rm A}}
\def\Brm{{\rm B}}
\def\crm{{\rm c}}
\def\Crm{{\rm C}}
\def\Drm{{\rm D}}
\def\Erm{{\rm E}}
\def\erm{{\rm e}}
\def\Frm{{\rm F}}
\def\grm{{\rm g}}
\def\Grm{{\rm G}}
\def\Hrm{{\rm H}}
\def\irm{{\rm i}}
\def\Irm{{\rm I}}
\def\krm{{\rm k}}
\def\Krm{{\rm K}}
\def\Lrm{{\rm L}}
\def\mrm{{\rm m}}
\def\Mrm{{\rm M}}
\def\Nrm{{\rm N}}
\def\nrm{{\rm n}}
\def\Orm{{\rm O}}
\def\orm{{\rm o}}
\def\Prm{{\rm P}}
\def\qrm{{\rm q}}
\def\Qrm{{\rm Q}}
\def\rmq{{\rm q}}
\def\Rrm{{\rm R}}
\def\srm{{\rm s}}
\def\Srm{{\rm S}}
\def\Trm{{\rm T}}
\def\trm{{\rm t}}
\def\Urm{{\rm U}}
\def\vrm{{\rm v}}
\def\Vrm{{\rm V}}
\def\Xrm{{\rm X}}
\def\xrm{{\rm x}}
\def\yrm{{\rm y}}
\def\zrm{{\rm z}}
\def\Zrm{{\rm Z}}


\def\hc{\kern 1pt {\ttsc} \kern -1pt}  
\def\hM{\kern 1pt {\ttsM} \kern -1pt}
\def\hN{\kern 1pt {\ttsN} \kern -1pt}
\def\hq{\kern 1pt {\ttsq} \kern -1pt}

\def\bbox{\vrule height5pt width5pt depth0pt}


\def\dag{\dagger}
\def\ddag{\ddagger}
\def\ox{\otimes}
\def\opl{\oplus}
\def\Oplus{\bigoplus}
\def\plusdotx{{\dot +}}
\def\rx{\stackrel{\rightarrow}{\times}}
\def\lx{\stackrel{\leftarrow}{\times}}
\def\rox{\stackrel{\rightarrow}{\ox}}
\def\lox{\stackrel{\leftarrow}{\ox}}
\def\from{\leftarrow}


\def\Amp{\mathop{\rm Amp}\nolimits}
\def\ACT{\mathop{\hbox{\rm A{\scriptsize{CT}}}}\nolimits}
\def\Acti{\mathop{\rm Act}\nolimits}

\def\Act{\mathop{\rm Act}\nolimits}
\def\Alg{\mathop{\rm Alg}\nolimits} 

\def\Amp{\mathop{\rm amp}\nolimits}
\def\AND{\mathop{\hbox{ \ts {\rm small AND} \ts}}\nolimits}
\def\apostrophe{\mathop{\hbox{\kern-3pt '\kern.5pt}}}
\def\ap{\mathop{\hbox{\kern-3pt '\kern.5pt}}}
\def\Arrow{\mathop{\rm Arrow}\nolimits}
\def\Arr{\mathop{\rm arr}\nolimits}
\def\Aut{\mathop{\rm aut}\nolimits}
\def\Av{\mathop{\rm av}\nolimits}
\def\Bas{\mathop{\rm bas}\nolimits}
\def\Cent{\mathop{\rm Cent}\nolimits} 
\def\Co{\mathop{\rm Co}\nolimits}
\def\Codeg{\mathop{\rm codeg}\nolimits}
\def\Col{\mathop{\rm col}\nolimits} 
\def\col{\mathop{\rm col}\nolimits}
\def\Const{\mathop{\rm Const}\nolimits}
\def\constant{\mathop{\rm constant}\nolimits}
\def\Coop{\mathop{\rm coop}\nolimits}
\def\Coord{\mathop{\rm coord}\nolimits}
\def\Cov{\mathop{\rm Cov}\nolimits} 
\def\Ctr{\mathop{\rm Ctr}\nolimits} 
\def\Def{\mathop{\rm Def}\nolimits}
\def\Deg{\mathop{\rm Deg}\nolimits}
\def\Det{\mathop{\rm Det}\nolimits}
\def\Diag{\mathop{\rm Diag}\nolimits}
\def\Dim{\mathop{\rm Dim}\nolimits}
\def\Diff{\mathop{\rm Diff}\nolimits}
\def\DIRAC{\mathop{\hbox{\rm D{\scriptsize{IRAC}}}}\nolimits}
\def\divm{\mathop{\rm div}\nolimits} 
\def\Endo{\mathop{\rm endo}\nolimits}
\def\EQ{\begin{equation}}
\def\ENDEQ{\end{equation}}
\def\Exp{\mathop{\rm Exp}\nolimits}
\def\Ext{\mathop{\rm ext}\nolimits}
\def\EXT{\mathop{\hbox{\Crm Ext}}\nolimits}
\def\EUCLID{\mathop{\hbox{\rm Euclid}}\nolimits}
\def\FALSE{\mathop{\hbox{\rm{F{\scriptsize{ALSE}}}}}\nolimits}
\def\Fin{\mathop{\rm Fin}\nolimits}
\def\GALILEO{\mathop{\hbox{\rm{G{\scriptsize{ALILEO}}}}}\nolimits}
\def\Gal{\mathop{\rm Galois}\nolimits}
\def\GL{\mathop{\rm GL}\nolimits}
\def\Grp{\mathop{\rm Grp}\nolimits} 
\def\hom{\mathop{\rm hom}\nolimits}
\def\Id{\mathop{\rm id}\nolimits}
\def\ISL{\mathop{\rm ISL}\nolimits} 
\def\im{\imath}
\def\IMPLIES{\mathop{\hbox{\rm{I{\scriptsize{MPLIES}}}}}\nolimits}

\def\Ind{\mathop{\rm ind}\nolimits}
\def\Inf{\mathop{\rm inf}\nolimits}
\def\In{\mathop{\rm In}\nolimits}

\def\Inv{\mathop{\rm Inv}\nolimits}
\def\IS{\mathop{\hbox{\rm{{\scriptsize{IS}}}}}\nolimits}

\def\Lie{\mathop{\rm Lie}\nolimits}
\def\Loc{\mathop{\rm loc}\nolimits}
\def\Lor{\mathop{\rm Lor}\nolimits}
\def\LORENTZ{\mathop{\hbox{\rm{L{\scriptsize{ORENTZ}}}}}\nolimits}
\def\Map{\mathop{\rm map}\nolimits}
\def\Matrix{\mathop{\rm Ma}\nolimits}
\def\Med{\mathop{\rm med}\nolimits}
\def\MINKOWSKI{\mathop{\hbox{\rm{M{\scriptsize{INKOWSKI}}}}}\nolimits}

\def\mod{\mathop{\rm mod}\nolimits}
\def\mRe{\mathop{\rm Re}\nolimits}  
\def\Net{\mathop{\rm Net}\nolimits}

\def\NOT{\mathop{\hbox{\rm{N{\scriptsize{OT}}}}}\nolimits}
\def\oblique{\mathop{\rm oblique}\nolimits}

\def\OP{\mathop{\hbox{\rm{o{\scriptsize{P}}}}}\nolimits}
\def\Op{\mathop{\rm Op}\nolimits}
\def\Orb{\mathop{\rm Orb}\nolimits}
\def\Ord{\mathop{\rm Ord}\nolimits}
\def\OR{\mathop{\hbox{\CSC\ts or\ts}}\nolimits}
\def\ox{\otimes}
\def\PAND{\mathop{\hbox{\CSC pand }}\nolimits}
\def\PG{\mathop{\rm PG}\nolimits}
\def\POINCARE{\mathop{\hbox{\rm{P{\scriptsize{OINCAR\'E}}}}}\nolimits}
\def\Point{\mathop{\hbox{\rm Point}}\nolimits}
\def\POR{\mathop{\hbox{\CSC por}}\nolimits}
\def\Prob{\mathop{\rm prob}\nolimits}
\def\Ray{\mathop{\rm ray}\nolimits}
\def\Scalar{\mathop{\rm Scalar}\nolimits}
\def\SDiff{\mathop{\rm SDiff}\nolimits}
\def\Seq{\mathop{\rm req}\nolimits}
\def\Ser{\mathop{\rm ser}\nolimits}
\def\Set{\mathop{\rm Set}\nolimits}
\def\SET{\mathop{\hbox{\CSC Set}}\nolimits}
\def\Sib{\mathop{\rm sib}\nolimits}
\def\Sim{\mathop{\rm sim}\nolimits}
\def\SL{\mathop{\rm SL}\nolimits}
\def\State{\mathop{\rm State}\nolimits}
\def\Sup{\mathop{\rm sup}\nolimits}
\def\Spin{\mathop{\hbox{\rm Spin}}\nolimits}
\def\Space{\mathop{\hbox{\rm Space}}\nolimits}
\def\Star{\mathop{\rm Star}\nolimits}
\def\SU{\mathop{\rm SU}\nolimits}
\def\Tan{\mathop{\rm Tan}\nolimits}
\def\Tensor{\mathop{\rm Tensor}\nolimits}
\def\Time{\mathop{\hbox{\CSC Time}}\nolimits}
\def\TRANSLATION{\mathop{\hbox{\rm{T{\scriptsize{RANSLATION}}}}}\nolimits}
\def\True{\mathop{\hbox{\CSC true}}\nolimits}
\def\Tr{\mathop{\rm tr}\nolimits}
\def\tr{\mathop{\rm tr}\nolimits}
\def\rmU{\mathop{\rm U}\nolimits}
\def\vac{\mathop{\rm vac}\nolimits}
\def\Vector{\mathop{\rm Vector}\nolimits}

\def\vertical{\mathop{\rm vertical}\nolimits}
\def\XAND{\mathop{\hbox{\CSC xand}}\nolimits}
\def\XOR{\mathop{\hbox{\CSC xor}}\nolimits}
\def\x{\times}


\def\arrow{\prec \kern -7pt - \kern-4.5pt \prec}
\def\arr{{\prec \kern -7pt -}}
\def\row{{ - \kern-4.5pt \prec }}
\def\rrawor{{ -\kern-7pt\succ \succ \kern -7pt -}}

\def\dar{\downarrow}
\def\Dar{\Downarrow}
\def\uar{\uparrow}


\newcommand{\cev}[1]{\stackrel{\leftarrow}{#1}}

\def\kkkk{\kern 4pt}

\def\bsl{\backslash}

\def\deq{\mathrel{\ol{:=}}}

\def\bu{\bullet}

\def\itemc{\item{$\circ$}}
\def\itemb{\item{$\bullet$}}
\def\parac{\noindent$\circ$\kern10pt }
\def\parab{\noindent$\bullet$\kern10pt}

\def\lhar{\leftharpoonup}
\def\lhas{\stackrel{\sim}{\lhar}}
\def\rhas{\stackrel{\sim}{\rhar}}
\def\rhar{\rightharpoonup}
\def\lharu{\leftharpoonup}
\def\rharu{\rightharpoonup}
\def\lhard{\leftharpoondownp}
\def\rhard{\rightharpoondown}


\def\Av{\mathop{\rm Av}\limits}

\def\ls{\vskip\baselineskip}
\def\lsn{\vskip\baselineskip\noindent}
\def\hls{\vskip0.5\baselineskip}
\def\hlsn{\vskip0.5\baselineskip\noindent}
\def\hhls{\vskip0.25\baselineskip}
\def\hhlsn{\vskip0.25\baselineskip\noindent}
\def\rls{\vskip-\baselineskip}
\def\rhls{\vskip-0.5\baselineskip}
\def\rhlsn{\vskip-0.5\baselineskip\noindent}
\def\rhhls{\vskip-0.25\baselineskip}
\def\rhhlsn{\vskip-0.25\baselineskip\noindent}
\def\nls{\noalign{\vskip\baselineskip}}
\def\nhls{\noalign{\vskip0.5\baselineskip}}
\def\nhhls{\noalign{\vskip0.25\baselineskip}}
\def\rnls{\noalign{\vskip-\baselineskip}}
\def\rnhls{\noalign{\vskip-0.5\baselineskip}}
\def\rnhhls{\noalign{\vskip-0.25\baselineskip}}

\def\onehalf{{^1\kern-1.5pt/\kern-1pt\sst 2}}
\def\abs{{\kern 1pt \vrule width.5pt height8pt depth3pt\kern 1pt}}
\def\square{\hbox{\hbox to 0pt{$\sqcup$\hss}\hbox{$\sqcap$}}}
\def\qed{\ifmmode\square\else\hfill\square\fi}
\def\qedm{\ifmmode\square\else\square\fi}

\def\qedh{\hfill\break\vskip0.01truemm\vskip-\baselineskip{\hfill\qed}}
\def\vqed{{\vskip-7truemm\hfill\qed}}
\def\vqeda{{\vskip-8.5truemm\hfill\qed}}
\def\vqedb{{\vskip-11.5truemm\hfill\qed}}

\def\cupp{{\cup\hskip-2truemm\cdot\hskip1.5truemm}}
\def\cuppd{\mathop{\cupp}}
\def\bcupp{\mathop{\bigcup{\hskip-3.5truemm\cdot\hskip2truemm}}}
\def\qoza{\mathop{q}}
\def\qoz{\qoza\limits}

\def\liminfuu{\hbox{\rm lim\ inf$\hskip0.5truemm$}}
\def\limsupuu{\hbox{\rm lim\ sup$\hskip0.5truemm$}}
\def\liminfu{\mathop{\vphantom{\tst\sum}\hbox{\liminfuu}}}
\def\limsupu{\mathop{\vphantom{\tst\sum}\hbox{\limsupuu}}}
\def\Maxuu{\hbox{\rm Max$\hskip0.5truemm$}}
\def\Minuu{\hbox{\rm Min$\hskip0.5truemm$}}
\def\Infuu{\hbox{\rm Inf$\hskip0.5truemm$}}
\def\Supuu{\hbox{\rm Sup$\hskip0.5truemm$}}
\def\Maxu{\mathop{\vphantom{\tst\sum}\hbox{\Maxuu}}}
\def\Minu{\mathop{\vphantom{\tst\sum}\hbox{\Minuu}}}
\def\Infu{\mathop{\vphantom{\tst\sum}\hbox{\Infuu}}}
\def\Supu{\mathop{\vphantom{\tst\sum}\hbox{\Supuu}}}

\def\dst{\displaystyle}
\def\hf{\hfill}
\def\sq{\sqrt}
\def\tst{\textstyle}
\def\sst{\scriptstyle}    
\def\ssst{\scriptscriptstyle}

\def\btd{{\tst\bigtriangledown}}
\def\ci{\circ}
\def\tom{\mapsto}

\def\mapsfrom{\leftarrow\kern -1pt
					\vrule height 5pt depth 0pt width 2pt}
\def\ltom{\longmapsto}
\def\Lar{\Leftarrow}
\def\Rar{\Rightarrow}
\def\LRa{\Leftrightarrow}
\def\da{\downarrow}
\def\ua{\uparrow}
\def\Llra{\Longleftrightarrow}
\def\lrar{\leftrightarrow}
\def\lra{\longrightarrow}
\def\Lra{\Longrightarrow}
\def\llra{\longleftrightarrow}
\def\lla{\longleftarrow}
\def\Lla{\Longleftarrow}
\def\hra{\hookrightarrow}
\def\hla{\hookleftarrow}
\def\un{{\infty}}
\def\bcu{\bigcup}
\def\Cap{\bigcap}
\def\Cup{\bigcup}
\def\es{\emptyset}
\def\pma{\pmatrix}
\def\pa{\partial}
\def\fa{\forall}

\def\sbq{\subseteq}
\def\sbp{{\subset\hskip-3truemm\cdot\hskip1.5truemm}} 
\def\sp{\supset}
\def\spq{\supseteq}
\def\we{\wedge}
\def\We{\bigwedge}
\def\Vee{\bigvee}
\def\ox{\otimes}
\def\Ox{\bigotimes}
\def\neqv{\not\equiv}

\def\ti{\times}
\def\sm{\setminus}
\def\intl{\int\limits}
\def\ointl{\oint\limits}
\def\ph{\phantom}
\def\phv{\vphantom}
\def\phh{\hphantom}
\def\wh{\widehat}
\def\wt{\widetilde}
\def\ov{\over}
\def\ol{\overline}
\def\ul{\underline}
\def\Bar{{\Big |}}
\def\bra{\langle}
\def\Bra{{\Big\bra}}
\def\ket{\rangle}
\def\Ket{{\Big\ket}}
\def\lst{\left/}
\def\rst{\right/}
\def\lr{\left(}
\def\rr{\right)}
\def\lst{\left|}
\def\rst{\right|}
\def\lss{\left\|}
\def\rss{\right\|}
\def\lc{\left\{}
\def\rc{\right\}}
\def\lb{\left[}
\def\lbb{\sqsubset}
\def\rb{\right]}
\def\rbb{\sqsupset}
\def\cl{\centerline}

\def\btwo{\hbox{\brm 2}}
\def\bthree{\hbox{\brm 3}}
\def\bfour{\hbox{\brm 4}}

\newdimen\linethickness  \linethickness=0.4pt
\newdimen\hboxitspacem    \hboxitspacem=2pt
\newdimen\vboxitspacem    \vboxitspacem=7pt
\newdimen\height
\newdimen\width
\def\boxitm#1{{%
   \setbox0=\hbox{$#1$}%
   \height=\dp0
   \advance\height by\vboxitspacem
   \advance\height by\linethickness
   \width=\wd0
   \advance\width by2\hboxitspacem
   \advance\width by2\linethickness
   \lower\height
   \rlap{\vbox{\hrule height\linethickness
               \hbox to\width{\advance\height by\ht0
                              \advance\height by-\linethickness
                              \advance\height by\vboxitspacem
                              \vrule height\height width\linethickness
                              \hfil
                              \vrule height\height width\linethickness}%
               \hrule height\linethickness}}%
   \kern\linethickness
   \kern\hboxitspacem
   \box0
   \kern\hboxitspacem
   \kern\linethickness}}

\newtheorem{prop}{$\circ$\kern-2pt}


\title{
Beneath Gauge}
\author{
David Ritz Finkelstein
\thanks{
School of Physics, Georgia Institute of Technology,
Atlanta Georgia 30332.
E-mail:
david.finkelstein@physics.gatech.edu
}
\and
Heinrich Saller
\thanks{
Heisenberg Institute of Theoretical Physics,
Munich, Germany D-80805.
E-mail: hns@dmumpiwh.bitnet
}
\and
Zhong Tang
\thanks{
School of Physics,
Georgia Institute of Technology,
Atlanta Georgia 30332.
E-mail:
zhong.tang@physics.gatech.edu.
Supported by the
M. \& H. Ferst Foundation.}
}
\maketitle

\begin{abstract}

Seeking a relativistic quantum infrastructure for gauge physics,
we analyze spacetime into three levels of quantum aggregation
analogous to atoms, bonds and crystals.
Quantum spacetime points with no extension
make up
more complex link units with
microscopic extension,
which make up networks
with macroscopic extension.
Such a multilevel quantum theory implies parastatistics for
the lowest level entities
without additional physical assumptions.
Any
hypercubical vacuum mode
with off-diagonal long-range order
that is covariant under $\POINCARE$ also has bonus
internal symmetries
somewhat like those of the standard model.
A vacuum made
with the dipole link proposed earlier
has too much symmetry.
A quadrupole link solves this problem.
\end{abstract}

\section {Quantum gauge theory}\label{sec:QGT}

Andrzej Trautman is a well known pioneer
in fiber bundle physics,
the invariant formulation of gauge theory.
We therefore focus here on the
quantum infrastructure of
classical gauge theories. We
dedicate these projections to him
in friendship and respect.

The fundamental problem of physics today
is to harmonize our spacetime and quantum concepts
and precepts.
Their discoverers, Einstein and Heisenberg,
violated each other's
main principles
and their theories suffered for it.
Einstein believed that the
Bohr-Heisenberg complementarity of descriptions
was less fundamental than the
alleged completeness of classical description.
His tangent vectors offend against
the Heisenberg uncertainty relation.
Heisenberg believed that
spacetime points
were not fundamental
and referred all quantities to one reference time.
His operator-algebra concept
infringes Einstein's locality principle;
as does the canonical quantization of Einstein's theory
in whatever variables.

This mutual neglect limits the meaningfulness (finiteness)
and validity of both their theories.
Present physics pays lip service to both locality
and complementarity
and conforms to neither.
To harmonize the two
we need to fundamentally rethink
them both;
to quantize spacetime and
localize quantum theory
both at once
and in a mathematically meaningful way.

Our project has drawn inspiration from
the adamantine ether of Isaac Newton,
the quantized spacetime of Hartland Snyder,
the spin networks of Roger Penrose,
the superconducting vacuum of Yoichiro Nambu,
and the solid spacetime of Andrei Sakharov.

The only known infrastructure for gauge theory
is the crystal underlying the Volterra theory of defects
and the Burgers vector field.\footnote
{\label{foot:Kleman}For these
see Kleman (1995).}
We therefore extend the atomic hypothesis
from chemistry to the rest of physics:
the world is a
crystalline network of linked
events.
The quantum spacetime ``atoms'',
quantum spacetime events without extension,
replacing the spacetime points
of present physics,
we call {\em topons}\/
(``present physics'' meaning especially general relativity
and quantum gauge field theory).
The quantum spacetime
linkages of topons having microscopic spacetime extension,
replacing the tetrads and tangent spaces
of present physics,
we call {\em chronons}\/.
A macroscopically extended quantum spacetime network itself,
replacing both the spacetime and fields of present physics,
we call a {\em plecton}\/.
A vacuum is
a fundamental mode of the plecton
that we use as a reference in describing excitations.
Just as molecular aggregates can have
approximately 0, 1, 2, or 3 space dimensions
(droplets, threads, films, pints)
and 0 or 1 time dimension
(brief or prolonged),
netwroks can have any dimensionality.
Newton made his vacuum crystalline
so that it might
support a transverse vibrational guide-wave
for polarized light-corpuscles.
Light is polarized just because
the electromagnetic field is a massless gauge
vector field.
Our leap from gauge to crystalline vacuum
merely updates
Newton's argument from polarization.

Newton did not say how
planets can fly through
his stiffer-than-diamond vacuum.
Now it is clear:  All
matter is
excitation of the vacuum plecton.
The vacuum is
a it supercrystal\/:
opaque to some of its excitations
and infinitely transparent to others,
as superfluids and superconductors
are opaque and transparent to theirs.
A supercrystalline Meissner-Higgs effect
expels excitations from ``good'' vacuum network,
concentrating them into spacetime
2-surfaces, strings.
All non-integrable gauge-transport arises
as in Volterra's theory of crystal defects
[outlined in Kleman (1995)]:
When we transport a gauge unit
of the spacetime supercrystal from one place to another,
its path threads between strings,
so the transport depends on the path.
These strings
manifest in present physics
as gauge flux tubes,
their ends
are particles.
Thus the particle gauge groups
directly characterize the causal microstructure of our spacetime
network,
and the particle masses and coupling constants
characterize the poles and residues in the
propagator of various excitations in the vacuum plecton.

The work goes so:

\S\ref{sec:BQND}  presents some
 qnd (quantum network dynamics) concepts and precepts.

\S\ref{sec:TOY}
is a toy qnd version
of elementary quantum mechanics
to explain the topon, chronon
and plecton by example.
The network dynamical equations follow from
independent variations of topons
in the plecton action.
Motion from one point of spacetime to another
is
a quantum jump or a sequence thereof.
We give no dynamics for a classical net.
Dynamics is a quantum phenomenon;
classical dynamics is a macroscopic quantum phenomenon.

\S\ref{sec:MI} presents a dipole vacuum plecton,
and its Poincar\'e and bonus symmetries.
Excitations of this vacuum can be corresponded to
gauge fields of the standard model and gravity.
This is a qnd analogue of a vector-field theory
of gravity. It seems
that the dipole network has no good four-dimensional vacuum
dynamics.
Worse, the dipole vacuum
does not sufficiently break quantum covariance.
It is invariant under the inhomogenous transformations
$\ISL(4, \Rbb)\supset\POINCARE$\/.
The dipole model seems defunct.

\S\ref{sec:MII}
at embarrassingly long last imitates in the quantum theory
of spacetime
the early evolution of Einstein from vector field to tensor in
the classical theory of spacetime.
The quadupole chronon is a directed pair of directed pairs
of topons.
The bonus symmetries and the correspondence principle
of the dipole model
survive the march to the quadrupole
unharmed.

\section{Basics of qnd}\label{sec:BQND}

\subsection{``All is quantum''}\label{sec:AIQ}
\subsubsection{Actors}
Quantum theory is a
semigroup theory of physical actions.
Kets and bras, though often called ``state vectors'',
represent sharp
modes of external action
that (as Ludwig says) produce or register quanta.
This action interpretation
is crucial for the alignment
of relativistic locality with quantum coherence,
and obviates any projection or ``collapse'' postulate.
To remember it we refer to
vectors that represent these actions as {\em actors}\/.

The
distinction between emission and absorption actions
ultimately
rests on the sign of an energy or frequency
and so cannot be exactly defined at one time $t$\/.
Truly time-local actors are
coherent superpositions of emission and absorption.

\subsubsection{Quantum principle}

{\em Each physical system
has a module $\calM$ and dual module $\calM^\dag$
of actors that projectively
represent sharp external actions for the system.
}

In particular the contraction $\bra m|l\ket$ vanishes
for forbidden transitions, and so represents a
transition amplitude.
This incorporates the operational interpretation
of quantum mechanics and
the irreducibility of its operator algebra.
It suggests that any classical
(central or super-selection) operators
are atavistic vestiges of classical physics.

\subsection{``Topology is all''}\label{sec:TIA}

{\em
There are spacetime points
and links among them.
The only dynamical variable is how the
points of spacetime are  linked.
Dynamical concepts and laws concern the near
neighbors of each point.}

This preserves
the monistic spirit of Einstein's unitary theories.
His metric tensor $g_{\n\m}$
specifies the directions and density of
causal connections, and nothing else.

\subsection {Topon}\label{sec:TOPON}

From \S\ref{sec:TIA} and \S\ref{sec:AIQ}
we infer a quantum spacetime point, the {\em topon}\/,
with a linear space (or module) $\calT$ of actors
replacing the point set of the spacetime manifold
of general relativity.
To make all
the higher-order topological elements quantum,
we assemble them from topons by quantum methods.

The actions of $\calT $
increment,
and those of the dual space ${\calT\,}^\dag$ decrement,
a grade that we will call topon number.

Since the state space of the classical spacetime point or event
actually varies from one spacetime model to another,
$\calT $  may eventually have to be
a quantum variable too (of a higher order).
Here we will
fix the actor space $\calT $
of the generic topon
as a given constant vector space
of dimension $|{\calT\,} |$\/.
The dimension $|{\calT\,} |$ is the number of possible distinct
spacetime events in the history of the system.
This is huge in any macroscopic experiment,
but possibly not in some small, high-energy, collision.
We sometimes take the limit $|{\calT\,} |\to\infty$\,.
Practically feasible actions,
however, directly affect a
number of spacetime points that remains finite though large
as $|{\calT\,} |\to\infty$\,.

\subsubsection{Indefinite metrics}

Hilbert space is
a finite-{} or infinite-dimensional actor space
with a positive-definite Hermitian form.
Today one recognizes that the physical Hermitian form
is determined by the dynamics
and is indefinite in theories
with Einsteinian locality [Saller (1996)]. We therefore
start without it. We work in an actor space with
the full general-linear invariance of projective geometry.
Hilbert space is an elegant trap.

\subsubsection{Quantum covariance}\label{sec:QC}
Then we infer from \S\ref{sec:TOPON} that qnd
concepts will be covariant and qnd laws
invariant under $\GL({\calT\,} )$.
We call this quantum covariance.
It replaces general covariance,
in the sense that a subgroup of $\GL({\calT\,})$
defined by the network ``condenses''
into the
Einstein invariance group of the spacetime manifold
as $|{\calT\,} |\to\infty$\ [Finkelstein (1996)].

Other subgroups of $\GL({\calT\,} )$ provide
other symmetries of present physics similarly.

\subsection{Chronon}\label{sec:CHRONON}

The spacetime of general relativity
is populated
by electromagnetic vectors, gravitational tensors,
and the like.
Therefore we suppose that the quantum spacetime
is an aggregate of linked topons, which
we call chronons or $\chi$\/, each characterized by
a fundamental time $\tbar$.

When (in a given basis)
a topon vector $\t$ appears as a factor in one chronon tensor
and its adjoint $\t^{\dag}$ in another,
the topon connects the two chronons
as a covalent bond connects two
atoms or a synapse connects two neurons.

In general, a chronon actor
$|X\ket$ is a tensor $X$
composed of
topon actors and duals,
regarded as a single vector.

In any model the chronon vectors constitute some module $\calX$\/
constructed from ${\calT\,}$ with dual module $\calX^\dag$\/.
Actions in $\calX$
increment a chronon number
and their duals in $\calX^{\dag}$ decrement it.

The main candidate chronons examined here are
\begin{itemize}
\item Dipole model:\quad $\calX={\calT\,} \ox
{\calT\,} ^{\dag}=[{\calT\,} \from{\calT\,} ]=:\calA$
\item Quadrupole model: \quad $\calX=\calA \ox
\calA ^{\dag}=[\calA\from\calA]=:\calB$
\end{itemize}

In the graph theory of spacetime,
spacetime is a set of arrows,
each joining two spacetime points.
This is a classical version of a dipole qnd.

The chronon actor space $\calX$
must condense into all the tensors
taken as fundamental in general relativity and the standard model.
Here we attempt this with but one kind of chronon;
but possibly the structure of the chronon
is a dynamical variable too.

\subsection{Plecton}

A quantum spacetime network or {\em plecton} in turn
is a network of chronons.
More formally: The local chronon sources $\chi\in \calX$
algebraically
generate a $\Zbb$-graded
huge-but-finite-dimensional
Grassmann ring $\calP$ over $\calX$.
This is the ring of plecton actors.
Its product $\vee$ is the concatenation of actions on the network.
Its grade is the chronon number of \S\ref{sec:CHRONON}.
We postulate a fixed structure for the chronon,
and the dimension of the plecton is an order parameter
determined by how chronons link up.

We designate by $\chi$ also the operator
$\chi\vee$ on $\calP$ of left
Grassmann multiplication by $\chi$.
We write $\chi^\dag:=\pa/\pa\chi$ for
Grassmann partial differentiation
with respect to $\chi$ from the left,
a dual actor,
expressed purely algebraically.

The meaning of the symbol $\chi^{\dag}$\/,
like any other partial derivative symbol,
depends on the basis.
Its $\dag$ is not a quantum-invariant adjoint operation.
But the module of
the operators $\chi^\dag$ (for all $\chi \in \calX$ )
has invariant meaning.

\subsubsection{Statistics and parastatistics}\label{sec:STATISTICS}

Using the Grassmann ring $\calP=\Vee\calX$ for plecton actors
assigns Fermi-Dirac statistics
to the chronon within the plecton.
We find three indications for this
and none to the contrary:

\begin{itemize}

\item The ensuing exclusion principle keeps
the universe from collapsing into a single
multiply-occupied
chronon actor
and maintains its cosmological size
at low temperature.

\item A classical manifold is a {\em set}
(say of tangent vectors and other tensors),
and the formal linearization of the power set of a set
is the
Grassmann algebra of an actor space.

\item The fermionic algebra, unlike the
bosonic,
is finite dimensional, so all its
operators are bounded.
This avoids mathematical nonsense.
\end{itemize}

Similarly we assign Fermi-Dirac statistics
to the topon within the chronon.
P. Gibbs (1995) also proposes
Fermi-Dirac statistics for spacetime points
(``event-symmetry'').
Here this implies an iterated Grassmann algebra
for the network:
$\calP=\Vee\calX=\Vee\Vee\calT$.

An iterated Grassmann algebra
implies parastatistics.
A product extensor in $\calP$
changes sign when two of its factors are interchanged,
but not necessarily
when two factors of its factors are interchanged.
This simple quantum fact corresponds to
a simple classical one:
Any  set (for example, $\{\{1,2\},\{3,4\}\}$) is
invariant under the interchange
of {\em its} elements (say, $\{1,2\}$ and $\{3,4\}$\/),
but not necessarily
under the interchange of {\em their} elements (say, 1 and 3).
Parastatistics is an inevitable consequence
of a deeper quantum theory, not a separate physical hypothesis.
Quark color,
introduced to avoid quark parastatistics,
may be a surface manifestation
of deep (multilevel) structure.

\subsubsection{Gauge}
By the {\em gauge element} we mean
the physical entity that
the basic gauge connection
connects.
For Weyl the gauge element was a length standard
like a machinist's
gauge block.
A {\em gauging} \/({\em Eichung}\/, calibration) is
a choice of a gauge element at each spacetime point.
For the gauge group $\Diff$\/ of general relativity,
the gauge element can be a tetrad of tangent vectors.

\subsection{Hyperdiamond}
One expects the theory to determine several vacua,
some degenerate.

The simplest plecton actor that is
exactly invariant under
a natural Poincar\'e group on the topon
is a hypercubical network
built from the four-dimensional harmonic oscillator algebra
in \S\ref{sec:MI}.
We call this structure
{\it hyperdiamond} in optimistic
homage to Newton's ether.
Such a quantum spacetimes has classical antecedents:
\hls
\parac  Newton said that space is a brain
(the ``sensorium of the deity'');
hyperdiamond is a mode of a quantum cellular automaton.
\hls
\parac Newton later suggested that space  is
a crystal
of corpuscles vibrating superluminally
about their equilibrium positions
(``aether'');
hyperdiamond is a quantum hypercubic lattice
of
quantum jumps to and fro
along quantum light cones.
\hls
\parac Feynman formulated a theory of a two-dimensional spacetime
using a two-dimensional checkerboard with one piece
in motion.
Hyperdiamond is a
Feynman checkerboard in four dimensions
instead of two,
filled entirely with pieces
instead of just one,
with null connections only,
and with quantum fluctuations in its structure.
\hls
\parac The classical counterpart of hyperdiamond
suggests a classical
bouncing-ball computer
of the kind that Fredkin and Toffoli (1982)
used to model reversible computation
and Fredkin (elsewhere) to model a classical spacetime.
\hls
\parac G. t'Hooft (1993) has used similarly null-directed
nets for classical spacetimes.

\hls
\parac The classical hypercubic lattice
underlying hyperdiamond is
a proper four-dimensional generalization
of the actual diamond crystallographic structure
(Smith 1995).

\hls
Hyperdiamond is $\POINCARE$ invariant,
unlike the classical hypercubic lattice.
That is,
a faithful representation of the group $\POINCARE$
upon the topons of hyperdiamond
leaves the hyperdiamond actor invariant.

Hyperdiamond (like many other candidates
for a vacuum plecton)
provides both
a gauge element and a gauge connection.
Its  gauge element is the unit cell of the vacuum.
Its connection is defined by the pattern
of links in the ``good'' network
joining each cell to immediate neighbours,
as in Burgers' theory of
defects in ordinary crystals.
Non-integrability arises when
a closed path links with tubes of ``bad'' network.

Hyperdiamond
has bonus
$\SU(2)$ and $\SU(3)$ symmetries
commuting with the Poincar\'e symmetries
(\S\ref{sec:BSHI}),
and tentatively
identified with isospin and color groups.

\subsection{Metaquanta and metastatistics}\label{sec:META}

If $V$ is the actor space of any quantum entity $\e$,
then $\Alg (V)$ and $\Vee V$,  linear spaces constructed from $V$
by multiplicative and additive processes,
are the actor spaces of
hypothetical new quantum entities,
called $\Arrow \e$ and $\Set \e$, of one higher unitization
order than $\e$.
These tensor-forming operations can be iterated,
resulting in still-higher-order tensors.
Here {\em order} counts nested  $|\dots\ket$-brackets
while {\em grade} (or rank) counts
side-by side $|\dots\ket$-brackets.
When we increase the order of a tensor,
we introduce new kinds of indices with greater ranges of values,
while when we increase the grade of a tensor
we merely attach more indices of the old kind.
Thus the familiar process of second ``quantization''
(``quantification" is an older and better term for it)
produces
a second-order quantum from a first-order one,
and a third quantification produces a third-order quantum.
We call
all such higher-order tensors collectively metatensors.
We use metatensors as actors for higher-order quanta,
or metaquanta.
We use the prefix
``meta-'' in general to mean
``having a higher order
of unitization'',
somewhat as in ``metalanguage'' and ``metalogic''.

In index-free notation one writes
metatensors by nesting brackets
as in \S\ref{sec:SYM} and in
equations (\ref{eq:pamBig},\/ {\ref{eq:xmBig}).
This corresponds to writing classical higher-order sets by
nesting braces ${\big\{}\{\dots\}{\big\}}$\/.
In index notation, metatensor indices
are generated inductively
within metatensor algebra
as nesting sets of sets of $\dots$ sets of simple indices
suitably ordered;
thus metatensors have {\em meta-indices}.

Fermi statistics holds for topons in the same chronon:
$|\t 1\ket\vee |\t 2\ket=-|\t 2\ket\vee |\t 1\ket$\/.
But some metatensors
do not change sign
under the exchange
$|\t 1\ket \leftrightarrow
|\t 2\ket$\/,
such as
$\Bar|\t 1\ket\vee |\t 3\ket\Ket\vee
\Bar |\t 2\ket\vee |\t4\ket \Ket$\/.
The usual concepts of momentum and statistics
apply to excitations of the plecton,
two orders above topons.
In general metaquanta have metastatistics [Finkelstein (1996)].
Like the subquarks proposed by Harari (1979) and
Shupe (1979), and the urs of Weizs\"acker(1981) and Jacob (1979),
the ultimate subquark, the spacetime point,
has no statistics in the ordinary sense.

We assign grade 1 to the arrow actors of $\calA$\/.
Then the actor space $\calP=\Vee\calA$
has an induced $\Zbb$-grade
(operator) $g$, the Grassmann degree,
even-valued for quasi-boson statistics,
odd-valued for fermion
statistics, called the chronon number.

\subsection{Quantum semigroup}
In qnd
we can hope to account for the (classical)
gauge groups of gravity and the
standard model from the symmetries of the quantum net
(\S\ref{sec:BSHI}).
In general,
symmetries of a classical structure form a classical semigroup
but symmetries of a quantum structure
naturally form a quantum semigroup,
called a q dynamics in Finkelstein (1972).
The hypothetical quantum group element
is called a {\em groupon}\/.
A quantum semigroup $G$ may be represented
by a vector space (or module) $B$
of actions on the groupon;
in a quantum philosophy
a parameter too is an action,
represented by an operator.
$B$ has two unital algebra or ring
products, a {\em serial} product
of actions on groupons and a {\em parallel} product of
groupons themselves.
Neither need commute.
We call such a structure $B$ a {\em double} algebra or ring.
Our {\em quantum group} is a quantum semigroup whose
double ring contains an
inverse action $\Inv$ for the serial product,
usually called the antipode.
Most quantum groupists assume a Hopf postulate,
that the two products commute
in a certain sense.
The most basic quantum examples violate this postulate.
It seems to be a classical vestige; we do not impose it.
For details and references see Finkelstein (1996).

\subsection{Local and remote descriptions}

In a general local action of the experimenter on the system,
a neighborhood of the episystem acts on one of the system.
In very low energy work,
both terminals of the action are surrounded
by good vacuum hypercrystal,
and by using this background we can
isomorph\footnote{Now that ``morph''
has become a verb, why not ``isomorph''?}
the action from its actual time $t$
to one fixed reference time $t=0$.
We call this transport of Hilbert spaces
the {\em quantum connection}\/.
In elementary quantum theory
we do not have independent Hilbert spaces for every time;
we represent actions by their isomorphs
at some reference time
by an implicit fixed, integrable quantum connection.
The usual quantum theory
thus rests on mobile
actions, abstracted from any one time
and free to act at any time whatever.
We call such a representation a {\em remote description}\/.

A {\em local description} deals with actions where they actually act.

The line between local and remote description
of actions
crosses the familiar one
between local and remote action.
Actions of either kind have descriptions of either kind.
We call a theory ultra-local if both the dynamics and the description
are local.

Functional quantum field theory seems ultra-local,
though at the cost of mathematical meaningfulness.
Qnd seeks to regularize functional quantum field theory,
by quantizing spacetime,
as Planck regularized thermal radiation by quantizing
energy.

\subsection{Correspondence}

{\em The basic classical variables of spacetime-continuum physics
are parameters of
coherent states of the quantum net.}

\subsection{Equivalence}

If $\G$ is any Lie group of $d$ dimensions,
we designate the corresponding gauge group
with $d$ arbitrary smooth functions
on a spacetime manifold $M$
by $\G^M$. We may locally identify $\Diff(M)$,
the diffeomorphisms of the spacetime manifold $M$\/,
with $\TRANSLATION^M$ in any chart of $M$\/.

A true harmonization of quantum theory
with general relativity and the standard model
must (we propose) save
all their symmetries:
the unitary of the quantum theory,
the external of general relativity,
and the internal of the standard model.

\subsubsection{Classical equivalence principle}
\label{sec:CEP}
{\em Dynamics is invariant
under the gauge groups of general relativity
and the standard model.}

The individual identities of spacetime
points and other gauge element are irrelevant;
only the pattern of gauge connections has
physical meaning.

One ordinarily satisfies this principle
by having the derivation $\pa$
enter only in a gauge-invariant combination
$D_\m=\pa_\m - A_\m$\/.
$D_\m$ at any one point can be obtained from its
vacuum value $\pa_\m$ by a local gauge transformation
changing the vector connection from 0 to $A_\m$\/.
In this sense all gauge forces are locally equivalent to
gauge transformations.
Since a Galilean acceleration $g$\/,
\EQ
x'^\m=x^\m + \frac{\scriptsize 1}{\scriptsize 2}
g\/{{\d^\m}_1} (x^0)^2\/,
\ENDEQ
is a special case of a
gauge transformation,
this equivalence principle
indeed generalizes Einstein's.
It suggests a corresponding
quantum equivalence principle for qnd:

\subsubsection{Quantum equivalence principle}
\label{sec:QEP}
{\em Dynamics is quantum invariant.}

By {\it quantum invariance} or {\it quantum covariance}
we mean invariance or covariance under $\GL({\calT\,} )$\/.
The quantum covariant concepts are all those defined by
purely quantum combinatorial processes like superposition
and tensor product,
without
benefit of metric. This invariance could be considered
to be implicit
in the ``All is quantum'' principle of \S\ref{sec:AIQ}.

The bras and kets of the Schwinger action principle
express a duality
between input and outtake. This distinction
depends on  a global time
and so is not quantum covariant.
We therefore replace it with the
more fundamental quantum-covariant duality
between system and episystem
as in the following example.

\section {Toy chronons}\label{sec:TOY}
We illustrate topons and chronons with a
one-dimensional toy,
a non-relativistic particle on a line.
\subsection{Quantum theory with remote description} \label{sec:QTRD}
Only propagation
through time $t$ is local.
The variable $x$ at any time
is represented by an operator on a single Hilbert space
$\calH$ \/.
Propagation along the $x$-axis is unretarded.
The  Hamiltonian and Lagrangian operators on $\calH$ are
\EQ
H={{p^2}\over{2m}}+V(x),\quad L={{p^2}\over{2m}}-V(x)\/,
\ENDEQ

We discretize the usual time, taking $t\in \Zbb$
to be the time in units of a fundamental time $\tbar$\/.
The transition amplitude for a general experiment
with initial actor $|\a\ket$, time delay $t$,
and final source bra $\bra\w|$ is then
\EQ
\label{eq:ADIRAC}
A=\bra \w| U \cdot U \cdot \dots \cdot U|\a\ket=\bra\w|U^t|\a\ket
\ENDEQ
where $U=e^{-iH\tbar}$ is the propagator for one time jump.

For arbitrary variations $\d x(t) $
of the variables $x(t)$\/,
according to Schwinger,
\EQ
\d A = i\bra\w|\d S|\a\ket + o(\tbar^2)
\ENDEQ
as $\tbar\to 0$\/,
where
\EQ
S=\sum L_t\tbar\/.
\ENDEQ

\subsection{Local quantum description}

We introduce a basis of formal time-actors
$|t\ket\in\calL=\dag\dag\Zbb_T$
labeled by $t\in\Zbb_T$, a formal time line
of $T$ instants.
We use these to define the topon space
${\calT\,}=\calL\ox\calH$; the actors
\EQ
| \a,0\ket:= |\a\ket\ox|0\ket,\quad\dots \quad
\bra \w,T|:= \bra T|\ox\bra\w|\/
\ENDEQ
in $\calT$ and $\calT^\dag$; the chronon actors
\EQ
U[t+1\from t]:= |t+1\ket\ox U\ox
 \bra  t|\/,\quad 1[t+1, t]:= |t+1\ket\ox 1\ox \bra t|
\ENDEQ
in $\calX={\calT\,}\ox{\calT\,}^\dag$\/;
an {\em experiment tensor}\/
\EQ
E=\bra\w,T|\vee 1[T-1, T-2]\vee\dots\vee|\a,0\ket
\ENDEQ
in $\calP:=\Vee \calX$\/; and the {\em dynamics tensor}
\EQ
D:= U[T\from T-1]\vee\dots\vee U[1\from 0]\/
\ENDEQ
in $\calP^\dag$\/.
$E$ and $D$ are now mutually dual metatensors.
This means that there is a unique Wick-style contraction
giving the transition amplitude, which we can write as
a vacuum-to-vacuum transition amplitude
\EQ
A=\bra \vac|ED|\vac\ket\/.
\ENDEQ
Under variations $\d$ about the actual dynamics
the amplitude variation has the form
\EQ
\d A =\bra \vac|E\d \tilde{S} D|\vac\ket\/.\label{eq:TOYDA}
\ENDEQ
and is stationary for fixed $|E\ket$\/.
The tilde on $\tilde{S}$ [adopted from Adler(1995)]
means that the usual factor
$i/\hbar$ has not been extracted from the generator.
The variations in the toy chronon $U$ are induced by variations
in its two terminal topons.

$E\vee D$ is a time-local description of the entire
process under study,
also encoding the topology of spacetime.
The experiment tensor
encodes non-maximal (``classical'')
information about our actions on the system.
The dynamics problem is to infer the
dynamics tensor $D$ that
gives transition amplitudes consistent with experiment.

\subsection{Quantum-covariant dynamics}

In a general quantum covariant qnd,
we assume, there are still mutually dual experiment and dynamics
tensors,
describing the actions of the episystem and the system
as far as each is relevant to the other.
The general transition amplitude is
now a {\em nullset-to-nullset} amplitude
\EQ
A= \bra 1|ED|1\ket\/. \label{eq:ADE}
\ENDEQ
There is still an action operator $\tilde{S}$\/.
Now the action principle
states that under all
independent anticommuting variations
$\d \tau$ and $\d \tau^{\dag}$
in every topon actor $\t$
and its dual $\t^{\dag}$,
the transition amplitude varies by
\EQ
\d A=\bra 1| E \d \tilde{S}  D|1\ket \label{eq:dAD}
\ENDEQ
and is stationary for fixed $E$\/.
Here $\tilde{S}$ is a fixed quantum-invariant operator
on $\calP$\/.

Now the equations of motion (\ref{eq:dAD})
are constraints
or subsidiary conditions on the dynamics tensor $D$\/,
not relations between operators at different times.

\section{Dipole qnd} \label{sec:MI}

Many spacetimes of general relativity
are defined by the causal relation between two variable points,
and a spacetime measure.
In classical set theory
a (small) relation is defined by its extension,
the set of all pairs
in the relation.
This led us to represent
a spacetime net (and a quantum-spacetime causal topology)
as a set of arrows
representing the relation of immediate causal successor.
This classical model in turn gave rise to dipole qnd.

\subsection{Network concepts}
Here are some simple classical and quantum network concepts
used to define both dipole and quadrupole qnd.
\subsubsection{Derived net}
(Of any partially ordered set or poset
$\{S, \le\}$\/.)
This is the net (a graph) $\d S$ consisting of all the
arrows joining points of $S$ to their immediate successors
in the partial order $\le$\/.

For example, $\d\Nbb$ and $\d\Nbb^4$ designate the derived nets
of the posets $\Nbb$ and $\Nbb^4$ (\S\ref{sec:SYM}).
\subsubsection{Adjoints}
If $S$ is any set, we write $S^\dag$ or $\dag S$
for the set of all
mappings: $S\to\Cbb$. Then $S^{\dag\dag}=\dag\dag S$\/,
the dual space
of $\dag S$,
is the {\em formal linearization} of $S$\/.
If $o$ is any object with state space $S$,
by the {\em formal quantization} of $o$\/,
written $\dag\dag o$ or $o^{\dag\dag}$,
we mean the hypothetical quantum entity
with actor space $S^{\dag\dag}$\/.

In all our theories
there is a natural
quantum-covariant metric $|\dots|^2$
and indefinite adjoint $\dag: \calX\to\calX^{\dag}$.
In the dipole model this is essentially unique:
\EQ
|\chi|^2:=  \bra \chi|\chi\ket := \tr \chi^2 \/,
\ENDEQ
the trace of the operator square  $\chi^2$\/.
This quantum-invariant adjoint $\dag$ converts an arrow creator
into an annihilator of the reversed arrow:
\EQ
|\tau_1\from \tau_0\ket^\dag = \bra \tau_0\from \tau_1|\/.
\ENDEQ
Without arrow reversal the adjoint would not be quantum-invariant.

\subsection{Hyperdiamond I}
\label{sec:HDI}
The dipole net has the topon actor space
${\calT}_I:={\dag\dag}\Nbb^4$\/, the chronon actor space
$\calX_I=\Alg({\calT}_I)$\/, and the plecton actor space
$\calP_I=\Vee\calX_I$\/.

The simplest possible locally finite candidate
for a vacuum actor $|\vac \,\Irm\ket\in\calP_I$
that reduces the diffeomorphism group
to a linear group
seems to be a hypercubic structure
based on the topon actor space
$\calT_I$\/,
\EQ
\label{eq:vacket}
|\vac\,\Irm\ket=
\Vee_{\m=1}^4 {\Big |} \sum_{m\in\Nbb^4}|m\ket\bra m+1_{\m} |
{\Big \ket}\/,
\label{eq:vac}
\ENDEQ
containing all the nearest-neighbor links of the
hypercube.\footnote{
Finkelstein (1996) had  $\Vee$
where (\ref{eq:vacket}) has  $\sum$\/.
We turned to (\ref{eq:vacket})
because its
Poincar\'e group acts consistently on all three levels
of aggregation,
and because it has off-diagonal long-range order.}
The coordinate $m\in\Nbb^4$ locates the tail of the chronon
and the coordinate $\m=1, 2, 3, 4$ gives
the direction of the chronon. For $1_{\m}$ see \S\ref{sec:SYM}.

\subsubsection{Poincar\'e invariance}

We define
general Poincar\'e transformations of topons
by the coherent state method,
using
discrete Poincar\'e symmetry transformations
of the hypercubic lattice
as infinitesimal generators.

For this construction we imbed
$\Nbb^4$ in an auxiliary Minkowski spacetime $M^4$\/.
$M^4$ is neither position space nor momentum space
but is used only during this construction.
We put the
four axes of $\Nbb^4$ along future null vectors of $M^4$,
creating an isotropic null coordinate system whose
Minkowski dual metric takes  the ``anti-Euclidean'' binary form
\EQ
\chi^{\n\m}=1-\d^{\n\m}= 0, 1. \label{eq:gnm}
\ENDEQ
Isotropic null coordinates are important for our coherent
state construction and for our model of the particle symmetries
for the following reason:
\hls
\noindent{\em The $4!$ symmetries of the directed hypercubic graph $\d\Nbb^4$
are Lorentz transformations, 12 proper and 12 improper.}
\hls
\noindent More accurately: are restrictions to $\d\Nbb^4\subset M^4$
of such transformations on $M^4$\/.

We cannot approximate $\POINCARE$ as closely as we like by
symmetries in $\Grp(\d\Nbb^4)$, but we can find an exact
isomorph of $\POINCARE$ in the algebra ${\dag\dag}\Grp(\d\Nbb^4)$.
First we
define operators  $\tilde{p}_\m$ and $x^\m$ on the topon space
$\calT :=\dag\dag\Nbb^4$
using finite symmetries of $\d \Nbb^4$\/.
Then we use these as infinitesimal generators
of $\Diff$\/ and {\em a fortiori} of $\POINCARE$.

We define the topon translation generator $\tilde{p}_{\m}$
to decrement $m$
 by $1_{\m}$\/:
\EQ
\tilde{p}_{\m}|m+1_{\m}\ket := |m\ket \/,
\label{eq:pam}
\ENDEQ
setting $|m\ket := 0$ if any component of $m$ is negative.
Thus the vacuum (\ref{eq:vac})
is $\Vee_{\m} |\tilde{p}_{\m}$.\/
To satisfy the canonical commutation relations
\EQ
\label{eq:CCR}
[\tilde{p}_{\m}, x^{\l}]=\d^{\l}_{\m}
\ENDEQ
we define the topon coordinate operators to increment $x^\m$ thus:
\EQ \label{eq:xm}
x^{\m} |n\ket
:= |n+1_{\m}\ket(n+1)\/.
\ENDEQ

We extend these operators naturally from topons to chronons
preserving the canonical commutation relations (\ref{eq:CCR}).
Any dipole chronon actor ${\Big |}\chi{\Big\ket}\in \calA$
 is the unitization of a linear operator $\chi$
on the topon actors $\calT $.
The natural extension of $\tilde{p}$ as infinitesimal generator
 from $\calT $ to $\calA$ is by the commutator
\EQ
\tilde{P}_{\m}{\Big |}\chi{\Big\ket}
:= {\Big |}[\tilde{p}_{\m}, \chi] {\Big\ket}\/.
\label{eq:pamBig}
\ENDEQ
To preserve the canonical commutation relations,
we must extend the coordinate operator from topons to chronons
barycentrically:
\EQ
X^{\m}{\Big |}\chi{\Big\ket}
:= {{1}\over{2}}{\Big |}x^{\m} \chi+\chi x^{\m} {\Big\ket}\/,
\label{eq:xmBig}
\ENDEQ

Because the chronon operators $\tilde{P}_{\m}$ commute, they
generate a translation group
$\Rbb^4$ that clearly fixes the candidate vacuum:
\EQ
\tilde{P}_{\m}|\vac\,\Irm\ket=0\/.
\ENDEQ

We now apply the coherent state method to rotations.
The natural Lorentz generator for the topon
in the given vacuum is
\EQ
\tilde{l}_{\m\l}=x_{[\m}\tilde{p}_{\l]}\/,
\label{eq:Lml}
\ENDEQ
implicitly using the metric tensor inverse to (\ref{eq:gnm})
to lower an index.
This guarantees the correct commutation relations
between $\tilde{l}$\/, $\tilde{p}$\/, and $x$\/.
In particular, $x$ and $\tilde{p}$ transform as vectors under
$\tilde{l}$\/.

We extend the infinitesimal $\tilde{l}$ from topons to chronons
naturally as a commutator:
\EQ
\tilde{J}_{\m\l}{\Big |}\chi{\Big\ket}
:= {\Big |}[\tilde{l}_{\m\l}, \chi] {\Big\ket}\/,
\label{eq:Jml}
\ENDEQ
We may decompose this chronon angular momentum $\tilde{J}$ into
\EQ
\tilde{J}_{\m\l}=\tilde{L}_{\m\l}+\tilde{S}_{\m\l}\/,
\ENDEQ
 an external or orbital angular momentum
\EQ
\tilde{L}_{\m\l}:=X_{[\m}\tilde{P}_{\l]}
\label{eq:LmlX}
\ENDEQ
for
the barycenter of the chronon
plus a spin or internal angular momentum $\tilde{S}_{\m\l}$
commuting with $X^{\m}$.
It is now straightforward that the candidate
vacuum has angular momentum zero:
\EQ
\tilde{J}_{\m\l}|\vac\,\Irm\ket=0
\ENDEQ

The operators $x$ and $\tilde{p}$ thus defined
generate a representation of $\Diff$\/,
broken by hyperdiamond I.
We see that hyperdiamond I is $\POINCARE$-invariant.
Unfortunately it is even
$\SL(4,\Rbb)$ invariant; we never used the
Minkowski metric in its construction.

We turn now to the conspicuous discrete symmetries that remain.

\subsection{Bonus symmetries of hyperdiamond}
\label{sec:BSHI}

We take for granted the concepts of
 {\em semidirect product} $A\rx B=C=B\lx A$
of any groups (or semigroups)
$A$ and $B$\/ with a {\em realization} of
$A$ on $B$;
and the natural extension of this concept from groups
to their algebras (could this concept be new?),
the {\em semitensor product} $A\rox B=C=B\lox A$
of any two
algebras $A$ and $B$ with a representation of $A$ on $B$.
\subsubsection {Symmetries of the directed hypercube}
The partial link-preserving
mappings of $\d \Nbb^4$  onto itself
form the semigroup of $\d \Nbb^4$, $\Grp(\d \Nbb^4)$\/.
Their formal linear combinations
form the  algebra of $ \d\Nbb^4$,
$\Alg(\d \Nbb^4)$\/,
which includes the group algebra
$\dag\dag(\Srm(4))$ of the symmetric group $\Srm(4)$
on the four axes as a tensor-product factor.

In the pre-dawn of group theory,
Galois semifactored $\Srm(4)$
(acting on the four roots of a quartic polynomial)
into
abelian groups according to
\EQ
\Srm(4)=\btwo\rx (\bthree \rx \bfour_2)\/.
\ENDEQ
Now the Klein group $\bfour_2$
acts on the four
vertices $1234$ neighboring the origin
of the tesseract $\d\btwo^4$\/,
as the three bi-cycles $(12)(34)\/$,
$(13)(24)\/$, $(14)(23)\/$,
and their product,
the identity $\Id$.
The four vertices form a regular tetrahedron 1234
in a spacelike three-dimensional hyperplane.
The bi-cycles are three spatial rotations of the tetrahedron
1234 through $\pi$ about orthogonal axes,
and the identity 1.

The $\bfour_2$ symmetries and the discrete
translational endomorphisms of $\d\Nbb^4$
are restrictions of the $\POINCARE$ transformations
of \S\ref{sec:HDI}.
If we extend the transformations
$\btwo$ and $\bthree$ from topons to chronons and plectons
in the standard way for finite discrete symmetries, we arrive
at familiar improper Lorentz transformations.
Quantum spacetime opens another way
to extend these finite transformations: as if they
were infinitesimal generators,
the method of coherent states.

If a classical object $o$ supports the cyclic group $\Zbb_N$
then the formal quantization $\dag\dag o$
supports the unitary group $\SU(N)$\/.

\noindent {\em Proof}\/.
Let $S$ be the state space, and
$A:=[S\from S]$ the arrow semigroup,
of a classical object $o$\/.
Then $\dag\dag S$ is the actor space
and $\calA:=\dag\dag A  = \dag\dag[S\from S]$ is the operator algebra
of the quantum entity $\dag\dag o$\/.
Now replace $o$ in this by $o^{(2)}=[o\from o]$\/,
the arrow on $o$, and its state space $S$, therefore,  by
$S^{(2)}=A$\/.
Then the hypothetical quantum entity $\e^{(2)}=\dag\dag o^{(2)}$\/
has as actor space the algebra $A$\/;
and as operator algebra, the double algebra
$\calB=\calA^{(2)}:=\dag\dag(A\from A)$.

Now specialize to $A=\Zbb_N$.
Then $B=\dag\dag(\Zbb_N\from \Zbb_N)\sim \GL(N,\Cbb)$\/.
 $\SU(N)$ is the maximal compact subgroup
of this algebra $B$\/.$\bbox$

For $N=2$,\/ $3$ we see that in this sense
$\btwo$ leads to $\SU(2)$ and $\bthree$ leads to $\SU(3)\/$.
We propose that these
are the color and isospin groups
of the standard model
[Selesnick (1995); Finkelstein (1996)].
There seems nothing else in nature that they could be.

There is no shortage of candidates for hypercharge $\Urm(1)$,
the most universal of the three quark charges.
That is the problem.

In the usual field theory, we replace  constant
group parameters of symmetry groups by
$x$-dependent ones to account for forces.
In this way the translation group yields gravity,
the spin group yields
torsion, $\Urm(1)$ and $\SU(2)$ yield the electroweak fields,
and $\SU(3)$ yields the strong.
Aside from the choice of hypercharge $\Urm(1)$, we accomplish
the same end in qnd
by suitably ``cracking'' the vacuum
[Finkelstein (1996)].
Loops surrounding the cracks then
have holonomy elements appropriate to all these forces.

This touches on a possibly crucial test
of qnd (or the standard model).
The Galois decomposition
$\Srm(4)=\btwo \lx \bthree \lx \bfour_2$ is only a semidirect
product, not a direct product.
The three semi-factor subgroups are not normal subgroups.
The three unitary groups of the standard model are
normal subgroups,
whose double algebras commute and co-commute.
It is therefore physically important
(though mathematically trivial) that
the three double subalgebras
$\dag\dag \btwo,\,\dag\dag \bthree,\,\dag\dag \bfour_2\subset\dag\dag\Srm(4)$
commute but do not co-commute with one another.
(Proof:
Underlying the semitensor product of algebras is a true
tensor product of vector spaces.
The operators on these factor spaces commute as usual.)

The serial product of the double algebra $\dag\dag B$
reflects only
the vector-space structure of $\dag\dag  A$
and has nothing to do with
the algebra product of $A$ or the group product
of $G$.
The parallel product reflects the non-commutative group
product in $\Srm(4)$,
and so is also non-commutative.
Thus the groupons of the quantum groups
$\dag\dag\btwo$ and $\dag\dag\bthree$ in $S(4)$ commute,
but their group parameters do not.
This departure from the standard model must have
physical consequences, which
may support or eliminate the hyperdiamond
choreography of the vacuum.

Despite its spacetime roots,
qnd has basic spin 1/2
entities.
 The arrow actors
$+f\vee e^\dag$  and $-f\vee e^\dag$
both point from $e$ to $f$.
They may therefore change sign under $2\pi$ rotation
without reversing their
direction in spacetime.
This sign is a quantum phase with no separate spacetime meaning.
The chronon actors in hyperdiamond,
like the Dirac spin operators $\g_\l$
of manifold-based physics, support a natural $\Spin(1,3)$
[Finkelstein (1996)].

\subsection{Quadrupole qnd}
\label{sec:ENDMI}

We seek an invariant action and dynamics tensor for a
dipole quantum net.
Every plecton operator is a polynomial in operators
of chronon creation and annihilation
defined thus.

Let  $B=\{|n\ket| n=1,\,\dots,|{\calT\,}|\}$
be  a basis for $\calT$\/.
Let $B^\dag$ be the reciprocal basis for the dual space
${\calT\,} ^{\dag}$\/.
These combine to
make a basis
$B\x B^{\dag}:=\{|n\from m\ket:=\Bar |n\ket\ox\bra m | \Ket|
 n,\,m=1,\,\dots,\,|{\calT\,}|\}$)
for the arrow algebra
$\calA$\/.
$\GL({\calT\,} )$ transformations of the $\calT $ basis induce
representations of
$\GL({\calT\,} )$ on the spaces ${\calT\,}^{\dag}$ and $\calA$\/.
From the basis vectors $|n\from m\ket$
we make arrow creators and annihilators
\EQ
c(n\from m):=|n\from m\ket\vee,\quad a(m \from n):=
\pa/\pa|n\from m\ket
\ENDEQ
using left progressive multiplication
and the left Grassmann derivative.
These are raw material for the following invariants.

Relative to the basis $B\subset \calT $\/, the operator
$c(n\from m) a(m \from n)$ (not summed over $n$ or $m$)
is an arrow number for any arrow $|n\from m\ket$\/.
It is not quantum
covariant
but only part of the quantum tensor $c(s\from r) a(m \from n)$.
The total number of chronons, however,
is the quantum-invariant operator,
\EQ
N(1):=\sum_{n, m}c(n\from m) a(m \from n) =: \tr\/ c a\/.
\ENDEQ
The trace $\tr$ acts only on the implicit topon indices,
not on the implicit plecton indices of these operators.
The higher-degree invariant
$
N(n):=\tr c^{n}  a^n
$
can be considered to count connected paths of $n$ arrows.

More generally let $[m\Leftarrow n]$ represent either the operator
 $c(n\from m)$
creating an arrow, or the cogredient operator $a(m \from n)$
annihilating the opposite arrow.
Let us represent a contraction by connecting arrows.
Then any loop
of the form
\EQ
L:=\quad|\Leftarrow \Leftarrow\dots\Leftarrow|\/,
\ENDEQ
is a scalar invariant,
the bars indicating where the loop closes.
Any quantum-invariant action of the dipole net is an
algebraic combination of such loop actions.

In field theory, an action that
is a non-trivial product $L_1 L_2$
of two spacetime integrals of functions of the fields
violates locality.
Similarly in qnd a product of two non-trivial loops like $L$
in the action violates locality.
This restricts us to single loops  and their linear combinations.

To form the dynamical equation we make independent
variations $\d |\t\ket$ and $\d \bra \t|$
in every topon $|\t\ket$ and $\bra \t|$.
This opens the loop at every possible link
into a chain from $|\t\ket$ to $\bra \t|$\/.
We cannot make a four-dimensional net of such chains,
but only one-dimensional nets.
This chronon has no good dynamics.

If the correspondent of an arrow is a vector field,
this failure is not surprising.
The corresponding field theory would be a vector theory of gravity,
which would violate the equivalence principle.

\section{Quadrupole chronon} \label{sec:MII}
We turn therefore to a more complex chronon, analogous to
a tensor field and a Wilson plaquette.
The  quadrupole chronon is a
double arrow
$\chi\sim[(\tau\from\tau)\from(\tau\from\tau)]$, with actor spaces
\EQ
\calT_{II}=\calT_I,\quad \calX_{II}:=\Alg(\Alg({\calT\,}_{II}))\/,
\ENDEQ
with four indices instead of two,
\EQ
\chi=\left((\chi^n{}_m{}_k{}^l)\right)\/,
\ENDEQ
and two natural products instead of one:
the {\em serial} product
\EQ
\chi(2)\cdot \chi(1):=(\chi^n{}_m{}_r{}^s(2)\chi^r{}_s{}_k{}^l{}(1))\/,
\ENDEQ
and the {\em parallel} product
\EQ
\chi(2)\circ\chi(1):=(\chi^n{}_s{}_r{}^l(2)\chi^r{}_m{}_k^s(1))\/.
\ENDEQ
There are two contractions, serial and parallel:
\EQ
\tr_{\orm}\chi=\chi^n{}_m{}_n{}^m\/, \quad \tr \chi = \chi^n{}_n{}_k{}^k\/
\ENDEQ
We keep the forms of equations (\ref{eq:ADE},\ref{eq:dAD}).
The action is still a closed (boundaryless) network of chronons,
no longer a mere loop but a higher-dimensional
network bag or reticule.
The variation $\d \tilde{S}$
tears off a chronon, opening the reticule,
and the action principle
(\ref{eq:dAD}) shows how the entire tensor $D$
is made by linking such opened reticules at their openings.
A candidate vacuum that breaks $\GL(4, \Rbb)$ down to
$\POINCARE$ is
\EQ
|\vac\, \Irm\Irm\ket:= \chi^{\n\m}|\tilde{p}_\n\ket \ox |\tilde{p}_\m \ket\/.
\ENDEQ
The left-hand factor $\tilde{p}$ refers to the head arrow
of the double arrow $\uparrow\from\uparrow$; the right-hand,
to the tail arrow.
Now
we must
make
a local action from these four-terminal chronons,
and fit it to such a vacuum.

\section{Summary}

Unlike discrete graphs,
quantum nets
can be locally finite and yet
exactly Poincar\'e-invariant.
The simplest net that supports
conservation of angular momentum and
energy-momentum is hyperdiamond,
a four-dimensional semi-infinite
hypercubical one.
The unit cell has bonus $\SU(2)$ and $\SU(3)$
symmetries arising from improper Lorentz symmetries.
Excitations of hyperdiamond support
the gauge groups of gravity, torsion and the standard
model in a classical limit.
It seems impossible to construct
a physical quantum-invariant action
or vacuum with
dipole chronons, and possible to do so with
with quadrupole chronons.

\section {Symbols}
\label{sec:SYM}

\noindent $\vee$ is the Grassmann progressive (exterior) product,
representing disjoint union,
with grade $g(\a)$\/.
 \hls
\noindent $\wedge$ is the Grassmann
regressive product, representing
exhaustive intersection, a Grassmann dual to $\vee$\/.
\hls
\noindent $\tilde{X} :=i X/\hbar$
\hls
\noindent $\dag X:=X^{\dag}:=$ (for any set or space $X$)
the module of morphisms $X\to \Cbb$
\hls
\noindent $| m \ket^{\dag}:=$ (for any basic vector $| m \ket$)
a reciprocal vector $\bra m|$ with $\bra n|m\ket=\d^n_m$\/.
\hls
\noindent $1_\m:=$ the unit vector along the $\m$ axis,
with components
$
1_{\m}{}^{\n}:= \d_{\m}^{\n}\,.
$
\hls
\noindent ${\Big  |}X {\Big\ket}$\/:
extensor (Grassmann element) of  grade 1
in all Grassmann products,
whatever the grade of the contents $X$.
\hls
\noindent $\Vee X:=$ Grassmann algebra over $X$\/.
\hls
\noindent ${\calT\,} :=$  actor space of the quantum event or topon.
\hls
\noindent $\calA:=\Alg({\calT\,}):=$ actot space
of the quantum arrow.
\hls
\noindent $\calB=\Alg(\calA):=$ operator algebra
of the quantum arrow.
\hls
\noindent $\calX:=$ actor space of the chronon.
\hls
\hls
\noindent $\calP:=\Vee\calX$ actor space of the plecton.
\hls
\noindent $\Nbb:=$ poset of the natural numbers ordered by
$\le\,$.
\hls
\noindent $\Nbb^4:=$ poset of
quartets $p_\m, q_\m, \dots$ of natural numbers, partially ordered
by $$p \le q \, \equiv\,  \forall \m \,|\,p_\m \le q_\m\,. $$
\hls

\section*{References}
\noindent Adler, S. (1995)
{\em Quaternionic Quantum Mechanics and Quantum Field Theory}. Oxford.
\hls
\noindent Davis, A-C. and R. Brandenberger (editors) (1995).
{\it Formation and Interaction of Topological Defects.}
Plenum Press, New York.
\hls
\noindent Finkelstein, D. (1972). Space-Time Code. III.
{\it Physical Review} {\bf D5},  2922 (1972)
\hls
\noindent Finkelstein, D. (1996).
{\it Quantum Relativity\/}. Springer, Heidelberg.
And references cited there.
\hls
\hls
\noindent Gibbs, P. (1995b). Symmetry in the topological phase of string theory.
Preprint PEG-05-95, hep-th/9504149
\hls
\noindent --- (1996). The small-scale structure of space time:
a bibliographical review. {\it Intern. J. Theor. Phys.} to be published.
\hls
\noindent Harari, H. (1979). {\em Physics Letters} {\bf 86B}, 83.
\hls
\noindent Jacob, P. (1979). Ph. D. Thesis, Tech. Univ. of Munich.
\hls
\noindent Kleman, M. (1995). The topological classification of defects. In
Davis and Brandenberger (1995).
\hls
\noindent Saller, H. (1996). {\em The analysis of
time-space translations in quantum fields.}
Preprint MPI-PhT/96-6.
Submitted for publication.
\hls
\noindent Selesnick, S. (1995). Gauge fields on the quantum net.
{\it J. Math. Phys.} {\bf 36}, 5465-5479.
\hls
\noindent Shupe, M. A. (1979). {\em Physics Letters} {\bf 86B}, 87.
\hls
\noindent Smith, Frank (Tony) (1995).
We thank Tony Smith for pointing this out.
\hls
\noindent t'Hooft, G. (1993) Preprint THU-93/26, gr-qc/9310026.
Dimensional reduction in quantum gravity.
\hls
\noindent Weizs\"acker, C. F. v. (1981). {\em Quantum Theory
and the Structures of Time and Space.} Vol. 4. Hanser. Munich.

\end{document}